\documentclass[twocolumn,amsmath,amssymb,floatfix,prb,epsf,showpacs]{revtex4}

\usepackage{bm}
\usepackage{graphics}
\usepackage{graphicx}
\begin{document}

\title{Tunneling between 2D electron layers with correlated disorder: anomalous sensitivity to
spin-orbit coupling}
\author{V. A. Zyuzin, E. G. Mishchenko, and M. E. Raikh}

\affiliation{Department of Physics, University of Utah, Salt Lake
City, UT 84112}


\begin{abstract}

Tunneling between two-dimensional electron layers with mutually
correlated disorder potentials is studied theoretically. Due to this
correlation, the diffusive eigenstates in different layers are
almost orthogonal to each other. As a result, a peak in the tunnel
$I$-$V$ characteristics shifts towards small bias, $V$.  If the
correlation in disorder potentials is complete, the peak position
and width are governed by the spin-orbit coupling in the layers;
this coupling lifts the orthogonality of the eigenstates.
Possibility to use inter-layer tunneling for experimental
determination of weak {\em intrinsic} spin-orbit splitting of the
Fermi surface is discussed.
\end{abstract}
\pacs{73.40Gk, 71.70.Ej, 72.25.Rb}
\maketitle

\section{Introduction}

Knowledge of spin-orbit (SO) splitting, $\Delta$, of energy spectrum
in 2D electronic systems is important for design of spintronic
devices in two respects. First, a number of proposed schemes
directly utilize the SO coupling for manipulating electron spin
polarization by means of creating
spatially inhomogeneous structures \cite{Khodas04}. Second, in
proposed schemes
that are not based on SO splitting, the latter limits
the device performance via a SO-induced
decoherence time \cite{zutic,loss}. Experimentally, large values of
SO splitting can be extracted from conventional measurements, such
as the beats of the Shubnikov-de Haas oscillations
\cite{dorozhkin87,luo90}. This, however, requires that $\Delta\tau
>1$, where $\tau$ is the electron scattering time. Experimental
determination of $\Delta$ in the opposite limit, $\Delta\tau <1$,
poses a considerable challenge. One has to look for physical effects
which are {\it anomalously} sensitive to the SO-coupling. An example
of such effect is the weak localization/anti-localization crossover
in magnetoresistance \cite{Koga02,Miller03}. Tunneling measurements
offer another possibility. Even when $\Delta\tau <1$, a structure
related to $\Delta$ manifests itself in the $I$-$V$ characteristics,
provided that the disorder is long-range, so that $\Delta\tau_{\rm
tr} >1$, where $\tau_{\rm tr}$ is the transport scattering time
\cite{Apalkov06}.

In 1993, Zheng and MacDonald \cite{ZM} made an observation that, in
the absence of the SO coupling, calculations of tunneling
conductance between two parallel electron layers with {\it
short-range} but {\it correlated} disorder potentials is analogous
to the calculation of conductance of a single layer with long-range
disorder. Formally, both calculations require  solution of the
equation for the vertex functions, obtained by a summation of ladder
diagrams. For a single layer, the vertex function has a pole at
frequency $\omega=i/\tau_{\rm tr}$, where
\begin{equation}
\frac{1}{\tau_{\rm tr}}=8\pi^2 \nu \int d{\bf q} S({\bf
q})\Bigl[1-\cos{\theta_{{\bf p},{\bf p+q}}} \Bigr] \delta(p^2-|{\bf
p}+{\bf q}|^2),
\end{equation}
where $\nu=m/2\pi$ is the 2D density of states (per spin) and
$S({\bf q})$ is the Fourier component of the correlator of the
intralayer disorder potential, $V({\bf r})$: $S({\bf q}) =\int d{\bf
r}~e^{-i{\bf qr}} \langle V({\bf r}) V(0) \rangle$. For interlayer
tunneling, the pole of the vertex function is at $\omega=i/\tau_0$,
where $\tau_0$ is defined as \cite{ZM}
\begin{equation}
\label{vertex_cor} \frac{1}{\tau_{\tiny{0}}}=8\pi^2 \nu \int d{\bf
q} \Bigl[S({\bf q})- S_{LR}({\bf q}) \Bigr]\delta(p^2-|{\bf p}+{\bf
q}|^2),
\end{equation}
where similar to the above, $S_{LR}({\bf q})$ is the Fourier
component of the cross-correlator $\langle V_L({\bf r}) V_R(0)
\rangle$ of the disorder potentials in the two layers. The physics
captured by Eq.~(\ref{vertex_cor}) is that despite strong scattering
in each layer, the {\it true eigenstates} in both layers are almost
identical when $V_L$ and $V_R$ are strongly correlated. Then the
pole at $-i\omega=1/\tau_0\ll 1/\tau$ reflects the fact that
eigenstates in two layers with energy difference $\gtrsim 1/\tau_0$
are {\em almost orthogonal}.

Basing on the above analogy, pointed out by Zheng and MacDonald,
one would anticipate  anomalous sensitivity of the tunneling current
between two layers with short-range correlated disorder to the SO
splitting in the layers.
In the present paper, we
will illustrate this anomalous sensitivity for a particular example of
tunneling between two identical quantum wells, to which electrons are
supplied by a $\delta$-layer of donors, located in the middle plane.
\begin{figure}[location=h]
 \centering
 \includegraphics[width=6cm] {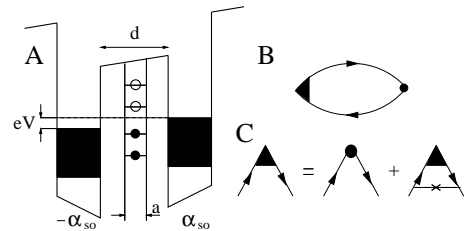}
 \caption{(A) Schematic illustration of a $\delta$-layer of donors located
symmetrically between two identical quantum wells. SO coupling
constants in the wells have equal magnitude and opposite signs; (B)
Diagram describing tunneling current between two wells with
correlated disorder; (C) Diagrammatic equation for the vertex
function.  }
 \end{figure}

\section{Tunneling current between two 2DEG with spin-orbit interaction}

The system under study is shown in Fig.~1. Once the donors get
ionized by yielding their electrons to the left and right electron
gases, electric fields which they create in both layers are equal in
magnitude and opposite in directions. As a result, the coupling
constants in the SO Hamiltonians \cite{SO} of the two layers are
{\it opposite}: $H^{(L)}_{\rm so} =\alpha_{\rm so} ({\bf p}\times
{\bm \sigma})_z$, $H^{(R)}_{\rm so} =-\alpha_{\rm so} ({\bf p}\times
{\bm \sigma})_z$. The important consequence of the geometry depicted
in Fig.~1 is that it allows to arrange correlation between {\it
spatial} wave functions in different layers \cite{drag}
corresponding to different energies separated by $2\Delta$. As a
result, $\Delta$ manifests itself in the
tunneling $I$-$V$ characteristics.

The tunneling Hamiltonian has the form,
\begin{equation}
\label{tunn_ham} H= t\sum_\alpha \int d^2r \left(\hat
\psi^{(L)\dagger}_\alpha({\bf r})  \hat \psi^{(R)}_\alpha ({\bf r})
+\hat \psi^{(R)\dagger}_\alpha({\bf r}) \hat \psi^{(L)}_\alpha ({\bf
r})\right),
\end{equation}
where $\hat \psi^{(L)}_\alpha({\bf r})$ and $\hat \psi^{(R)}_\alpha
({\bf r})$ are the electron operators in the two layers, and
$\alpha$ is the spin index. The overlap integral $t$ for the
size-quantization wavefunctions in the two layers is assumed to be
real, for simplicity. The tunneling described by
Eq.~(\ref{tunn_ham}) preserves both electron spin and momentum.

 Calculation of the interlayer tunneling current (see Fig.~1) reduces to
finding the vertex function for the case when  the electron Green's
functions in the layers are matrices. Namely, the retarded Green's
functions are
\begin{eqnarray}
\label{retards} \hat G^{(L)}_R(\epsilon,{\bf p}) =[\epsilon-\xi
-\alpha_{\rm so} ({\bf
p}\times {\bm \sigma})_z+i/2\tau]^{-1}, \nonumber\\
\hat G^{(R)}_R(\epsilon,{\bf p}) =[\epsilon-\xi +\alpha_{\rm so}
({\bf p}\times {\bm \sigma})_z+i/2\tau]^{-1},
\end{eqnarray}
where $\xi$ is the electron energy, measured from the Fermi level.
Advanced Green's functions are obtained from Eqs.~(\ref{retards}) by
reversing the sign of $i/2\tau$-terms. Solving the matrix equation
illustrated in Fig.~1 yields the following generalized expression
for the vertex
\begin{equation}
\label{solvert}  {\cal T} (\omega)= t
\frac{(\omega+i/\tau)^2-4\Delta^2}{(\omega+i/\tau)(\omega+i/\tau_0)-4\Delta^2},
\end{equation}
where $\Delta=\alpha p_F$, and $p_F$ is the Fermi momentum. In the
absence of the SO coupling Eq.~(\ref{solvert}) reduces to the result
${\cal T} (\omega)= t(1-i\omega\tau)/(1-i\omega\tau_0)$ of
Ref.~\onlinecite{ZM}. Incorporating the vertex function Eq.
(\ref{solvert}) into the standard expression~\cite{Mahan} for the
tunneling current
\begin{eqnarray}
\label{loop} I(V) &=& e^2 A t V   ~\Im \Bigl\{ \Gamma (eV)\nonumber\\
&& \times  \text{Tr} \int d{\bf p}~
 {\hat G}^{(L)}_R(0,{\bf p}) {\hat G}^{(R)}_A (-eV,{\bf
p}) \Bigr\},
\end{eqnarray}
we arrive to the final result,
\begin{equation}
\label{result}  I(V)= \frac{2e^2t^2 A\nu
V[4\Delta^2\tau^{-1}+(e^2V^2+\tau^{-2})\tau_0^{-1}]}
{[e^2V^2-4\Delta^2-\tau^{-1}\tau_0^{-1}]^2+e^2V^2(\tau^{-1}+\tau^{-1}_0)^2}.
\end{equation}
Here $A$ is the lateral area.

Anomalous sensitivity of the $I$-$V$ characteristics (\ref{result})
to the SO splitting is illustrated in Figs.~2-4. For large
splitting, $\Delta\tau > 1$ (see Fig.~2), correlation between the
disorder potentials is not important. The peaks  of the $I$-$V$
curves are located at $eV=\pm 2\Delta$, while the peak widths are
$\approx 1/\tau$. Such a form of the $I$-$V$ curves reflects the
fact that, with opposite signs the SO coupling constants in the
layers, the intralayer {\em spinor} eigenfunctions are maximally
correlated, when their energies differ by $2\Delta$.
\begin{figure}[location=h]
 \centering
 \includegraphics[angle=270, width=8.5cm] {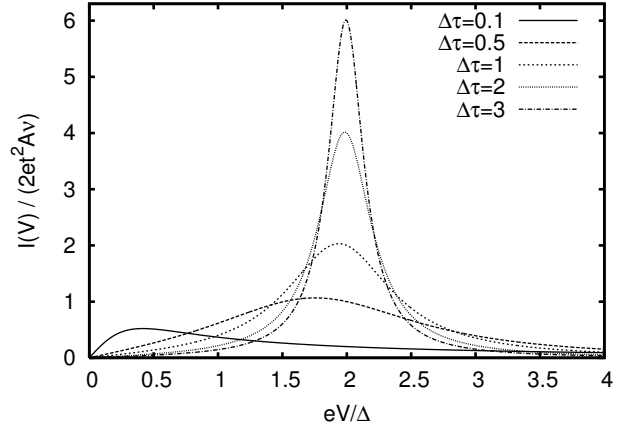}
 \caption{The tunnel $I$-$V$ characteristics is plotted from Eq. (5)
for different values of dimensionless scattering rate $\Delta\tau$ in the layers.
For $\Delta\tau >0.5$ the current is maximal at bias $eV=2\Delta$.}
 \end{figure}

As it is seen from Fig.~2, the position of the current maximum
rapidly shifts from $eV=2\Delta$ towards smaller biases for
$\Delta\tau \lesssim 0.5$. The $I$-$V$ characteristics for this case
are shown in Fig.~3. A remarkable feature of the curves in Fig.~3 is
their strong sensitivity to $\Delta$ when the disorder is strong,
$\Delta\tau < 1$, so that the characteristics  of {\em individual}
layers are {\em insensitive} to the SO coupling.
\begin{figure}[location=h]
 \centering
 \includegraphics[angle=270, width=8.5cm] {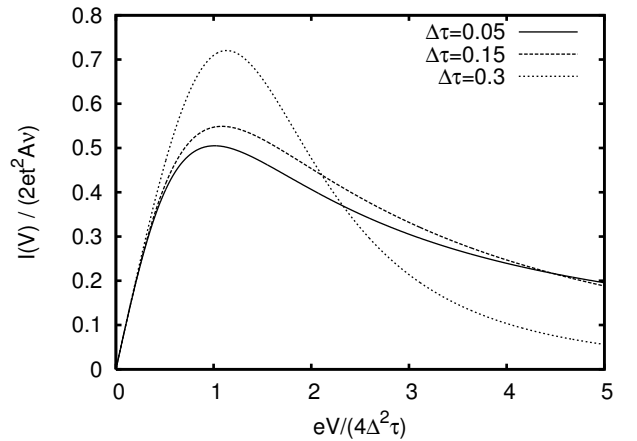}
 \caption{The tunnel $I$-$V$ characteristics is plotted from Eq. (5)
for fully correlated disorder potentials in the layers at different values of  $\Delta\tau \ll 1$.
 The current is maximal at $eV=4\Delta^2\tau$.}
 \end{figure}
For fully correlated disorder potentials in the layers,
$\tau_0\rightarrow \infty$, the $I$-$V$ characteristics for
different values of $\Delta\tau < 0.2$ fall on top of each other
when plotted as a function of the ratio $eV/4\Delta^2\tau$. Thus,
the position of maximum of the tunneling current allows to extract
the Dyakonov-Perel \cite{DP} spin decoherence time
$\tau_s=(2\Delta^2\tau)^{-1}$. The underlying reason is that, due to
opposite signs of the intralayer SO coupling constants, the
eigenfunctions in the layers are not orthogonal even if disorders
are fully correlated. Then $\tau_s^{-1}$ is a quantitative measure
of the energy interval in which the orthogonality is lifted. The
fact that position of the maximum in Fig.~(3) is at $eV=2/\tau_s$
reflects that electrons in {\it both} layers undergo spin
relaxation.

Incomplete correlation of disorder potentials, $V_L({\bf r})$ and
$V_R({\bf r})$, in the layers is another source of lifting of
orthogonality of eigenstates. This mechanism is quantified by the
energy scale $1/\tau_0$, defined by Eq.~(\ref{vertex_cor}). It might
be expected that in the presence of both mechanisms the maximum is
located at $eV=2/\tau_s+1/\tau_0$, which has a meaning of a combined
dephasing time. This is indeed the case, as illustrated in Fig.~4.


\section{Effect of electron-electron interactions}

Let us address the question whether the above SO-induced peaks
survive the presence of electron-electron interactions. Interactions
cause a {\em dynamic} lifting of orthogonality of the eigenstates,
and might result in the broadening of the peaks. We now demonstrate
that at zero temperature the peaks are robust, but eventually get
smeared away as the temperature increases.

\begin{figure}[location=h]
 \centering
 \includegraphics[angle=270, width=8.5cm] {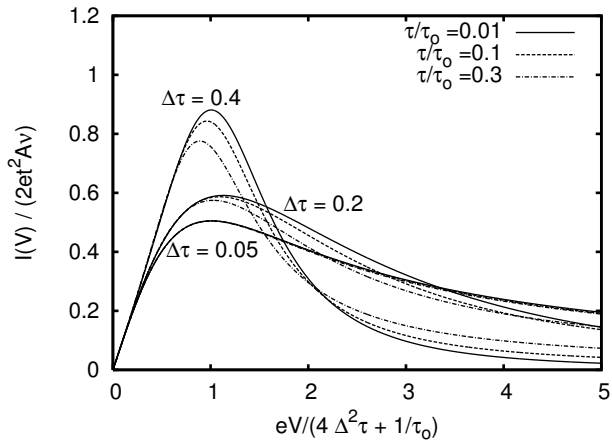}
 \caption{ $I$-$V$ curves for different values of spin-orbit coupling and different
 degrees of correlation of
 disorder in the layers. The maxima are at biases determined by the combined decoherence rate.
 The latter is dominated by $1/\tau_0$ term for $\Delta \tau =0.05$
 and by spin relaxation term, $4\Delta^2\tau$, for $\Delta \tau =0.4$.
 }
 \end{figure}

On the quantitative level, in order to incorporate both the
interactions and the correlated disorder into the theory, it is
convenient to express the tunneling current in terms of the {\it
exact} eigenfunctions, which are the same in the two layers. Let us
denote with $\psi_m({\bf r})$ the $m$-th eigenstate for a given
realization of disorder potential. The energy of this state is equal
to $\epsilon_m$ as electron-electron interactions are neglected. In
the presence of electron-electron interactions the retarded electron
Green function can be written as,
\begin{equation}
G_R(\epsilon,{\bf r}_1,{\bf r}_2) =\sum_m \frac{\psi_m({\bf
r}_1)\psi_m^\dagger ({\bf r}_2)}{\epsilon-\epsilon_m-\Sigma_m
(\epsilon)},
\end{equation}
where  $\Sigma_m (\epsilon)$ denotes the electron self-energy of the
$m$-th eigenstate.

The knowledge of the eigenfunctions suffices to evaluate the
tunneling current (in the lowest order in $t$) in a general form,
\begin{eqnarray}
\label{prior} I(V)=2et^2 \int\frac{d\epsilon}{2\pi}
\Bigl[n(\epsilon)-n(\epsilon+eV) \Bigr]\nonumber\\ \times \int d{\bf
r}_1d{\bf r}_2  {\cal A}^{(L)} (\epsilon, {\bf r}_1,{\bf r}_2){\cal
A}^{(R)} (\epsilon+eV, {\bf r}_2,{\bf r}_1),
\end{eqnarray}
and express it via  the Fermi-Dirac distribution $n(\epsilon)$ and
the {\it non-averaged} spectral functions in the left and right
layers, ${\cal A}^{(L)} (\epsilon, {\bf r}_1,{\bf r}_2)$ and ${\cal
A}^{(R)} (\epsilon, {\bf r}_1,{\bf r}_2)$, respectively. The
spectral function is determined by the difference of the retarded
and advanced functions in each layer,
\begin{equation}
{\cal A} (\epsilon, {\bf r}_1,{\bf r}_2) =\frac{i}{2} \Bigr[G_R
(\epsilon, {\bf r}_1,{\bf r}_2)-G_A(\epsilon, {\bf r}_1,{\bf r}_2)
\Bigl].
\end{equation}
At this point, we emphasize that, for fully correlated disorder
(neglecting spin-orbit interaction), we can perform coordinate
integration in Eq.~(\ref{prior}) prior to performing the
configuration averaging, and cast it in the form
\begin{eqnarray}
\label{prior1} I(V)=2et^2  \sum_m
\int\limits_{-eV}^0\frac{d\epsilon}{2\pi} ~  \Im
\left[\frac{1}{\epsilon-\epsilon_m-\Sigma_m
(\epsilon)}\right]\nonumber\\ \times \Im \left[\frac{1}{\epsilon
+eV-\epsilon_m-\Sigma_m (\epsilon+eV)}\right].
\end{eqnarray}
   The reason why
explicit integrations over ${\bf r}_1$ and ${\bf r}_2$ can be
performed in Eq. (\ref{prior}), leading to Eq. (\ref{prior1}),
is the mutual orthogonality of the eigenstates in the two layers
resulting from the fully correlated disorder. In Eq.~(\ref{prior1})
we  have also set $T=0$ in the difference of the Fermi functions.
This is justified as long as $T$ is much smaller than $E_F$. The
reason is that, similarly to the in-plane conduction, the
temperature dependence of the tunnel current comes exclusively from
the $T$-dependence of the self-energy, i.e. from inelastic
processes.

The question whether $\Sigma_m (\epsilon)$ and
$\Sigma_m(\epsilon+eV)$ in Eq.~(\ref{prior1}) can be replaced by the
disorder averaged values is highly non-trivial in the limit $T \to
0$. However, we can rigorously address the issue of smearing of the
SO-related peak in the $I$-$V$ curve for disordered layer by
treating interactions at the {\it perturbative} level, which
corresponds to the expansion of Eq.~(\ref{prior1}) to the first
order in $\Sigma_m$. This expansion yields,
\begin{eqnarray}
\label{prior2} \delta I(V)=\frac{2t^2\nu A}{e V^2}
\int\limits_0^{eV} d\epsilon
\Bigl[\gamma_\epsilon(\epsilon+eV)+\gamma_{\epsilon+eV}(\epsilon)
\Bigr],
\end{eqnarray}
where $\gamma_\epsilon(\omega) = \nu^{-1} \sum_m \langle
\delta(\epsilon-\epsilon_m)~ \Im \Sigma_m(\omega) \rangle$ is now
the disorder-averaged inverse inelastic lifetime.

(i) For $\Delta \tau \ll 1$ the peak position, $eV=4\Delta^2\tau$,
is below the elastic scattering rate, {\it i.e.}~at the energy
corresponding to the peak position the motion of electrons is
diffusive. The corresponding lifetime was studied in the seminal
paper Ref.~\onlinecite{Abrahams}, and was shown to be,
$\gamma(\epsilon) \sim |\epsilon|/E_F\tau$. We can now compare
Eq.~(\ref{prior2})  with the ``non-interacting'' value of current,
given by Eq.~(\ref{result}), at the bias corresponding to the peak
position, $eV= 4\Delta^2\tau$. We find that the  ratio $\delta I/I
\sim 1/E_F\tau$ is small {\it regardless} of the actual value of the
decoherence rate.

(ii) Similarly, for large values of spin-orbit splitting, $\Delta
\tau \gg 1$, we should utilize ballistic inverse lifetime,
$\gamma(\epsilon) = (\epsilon^2/4\pi E_F) \ln{[E_F/\epsilon]}$,
established in Refs.~\onlinecite{MJ,ZDS}. Comparison of
Eq.~(\ref{prior2}) with the value given by Eq.~(\ref{result}) at the
 peak position, $eV= 2\Delta$, we conclude
that the corresponding ratio is again small, $\delta I/I \sim
(1/E_F\tau)\ln{[E_F/\Delta]}$.

This suggests that, at zero temperature, interactions do not destroy
the SO-induced peak in the $I$-$V$ curve. However, this destruction
eventually happens upon increasing $T$. A crude estimate for the
temperature at which the peak is washed out by interactions can be
obtained by equating the peak position $eV=4\Delta^2\tau$ to
$\gamma(\epsilon=T)$. With logarithmic accuracy, this yields the
restriction $T < (\Delta\tau)^2 E_F$, so that even with $\Delta\tau
<1$ the peak survives at reasonably high temperatures.

To trace quantitatively the smearing of the peak with $T$, we first
note that ``single-electron'' $I$-$V$ characteristics (\ref{result})
can be obtained from Eq.~(\ref{prior1}) upon inserting spin
decoherence rate into the self-energy, $\Sigma_m \to 2i \Delta^2
\tau$, and replacing the sum over $\epsilon_m \to \xi $ by the
integral, $\sum_m \to \nu A \int d\xi$. As the next step, we take
into account the finite-$T$ decoherence by writing $\Sigma = 2i
\Delta^2 \tau+i\gamma_T$, where $\gamma_T=(T/2E_F\tau) \ln{(T_1/T)}$
\cite{Abrahams,Blanter} and $T_1=r_s^2E_F^4\tau^3$; here $r_s$ is
the interaction parameter of 2D electron gas. To utilize the
energy-independent $\gamma(T)$ in the self-energy, the temperature
must be large compared to the peak position, $eV=2\Delta^2\tau$.
This requirement does not contradict the restriction on the smearing
obtained from the above crude estimate. Indeed, both conditions can
be conveniently rewritten as, $T/E_F\tau \ll \Delta^2\tau \ll T$, so
that it is the large value of $E_F\tau$ which makes them consistent.
Upon the suggested replacements, the temperature-dependent $I$-$V$
characteristics follows from Eq.~(\ref{result}) with the spin
relaxation rate modifies as $2\Delta^2\tau \to
2\Delta^2\tau+\gamma_T$, with $\tau_0 =\infty$
\begin{equation}
\label{resultf}  I(V)= 4e^2t^2 A\nu \frac{ V(2\Delta^2\tau+\gamma_T)}
{e^2V^2+4(2\Delta^2\tau+\gamma_T)^2}.
\end{equation}
Eq. (\ref{resultf}) indicates that the position of maximum of the
$I$-$V$ curve shifts almost linearly with temperature, see Fig.~5.
This suggests that the SO relaxation rate can be inferred from
experiment even when measurements are performed at $T>
(\Delta\tau)^2E_F$. One has to plot the peak position as a function
of $T$ and extrapolate the data to $T\rightarrow 0$. Besides, the
actual restriction on  $T$ is  ``softer'' than the one obtained from
the crude estimate, by virtue of numerical coefficients in
$\gamma_T$ and spin relaxation time $1/\tau_s$. Indeed, the
requirement $\gamma_T \tau_s/2 <1$ imposes (neglecting logarithmic
factor) the following restriction, $T<4(\Delta\tau)^2E_F$.
\begin{figure}[location=h]
 \centering
 \includegraphics[angle=0, width=8.5cm] {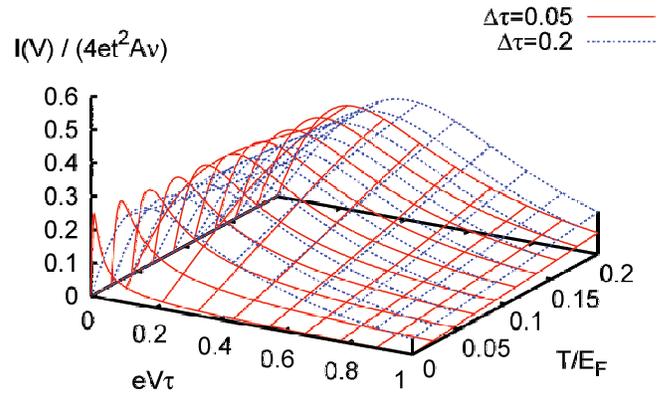}
 \caption{(Color online) Temperature dependence of the tunneling characteristics for different values of spin-orbit
 coupling constant and fully correlated disorder, from Eq.~(\ref{resultf}). Increasing $T$ results in the
broadening of the peak and its net shift towards
 higher biases.}
 \end{figure}

\section{Summary and Conclusions}

Our main finding is that, with correlated disorder in the layers,
the SO coupling causes a zero in $dI/dV$ even for $\Delta\tau \ll
1$, when the spin subbands in the layers are not resolved. For clean
layers with $\Delta\tau > 1$, sensitivity of tunneling current to
the SO coupling was  pointed out in Ref.~\onlinecite{RaichevDebray}.

The condition that position of zero in $dI/dV$ is due to the SO
coupling is that the contribution, $4\Delta^2\tau$, to the combined
decoherence rate exceeds $1/\tau_0$, caused by incomplete
correlation of disorders in the layers. To estimate the feasibility
to meet this condition, we assume that the origin of incomplete
correlation is a finite width, $a$, of the $\delta$-layer, see
Fig.~1. Assuming that the in-plane positions of donors with
concentration, $N_d$, are completely random, the Fourier components
of the correlators, entering the expression Eq.~(\ref{vertex_cor})
for $\tau_0^{-1}$, can be presented as $S_{LR}({\bf q})=
N_{d}|U({\bf q})|^2 e^{-qd}$; $S({\bf q})=N_{d}|U({\bf q})|^2
\sinh{(qa)}~e^{-qd}/qa$, where $d$ is the barrier thickness, and
$U({\bf q})$ is the Fourier transform of the potential created by a
donor in the layer. Then Eq.~(\ref{vertex_cor}) takes the form
\begin{eqnarray}
\label{last}
\frac{1}{\tau_0} = 2\pi\nu N_d \int d{\bf q}~|U({\bf q})|^2
\left(\frac{\sinh{qa}}{qa}-1\right) ~e^{-qd}.
\end{eqnarray}
Assuming that the screening radius in the layers is smaller than
$d$, we can set $U(1/d)\approx U(0)$. Then Eq.~(\ref{last}) yields
$\tau_0/\tau =(a/d)^2$, so that the condition $4\Delta^2\tau
>\tau_0^{-1}$ reduces to $\Delta\tau >a/2d$, i.e.\ the value
$\Delta\tau =0.05$, used in the numerics above, is quite feasible
for the atomically sharp $\delta$-layer.

As a final remark, the assumption, crucial for our calculations, was
that the barrier is spatially homogeneous, so that the tunneling
occurs with the conservation of the in-plane momentum. We had also
assumed that the positions of donors in the $\delta$-layer are
random. This randomness might, in principle, lift the momentum
conservation. The condition that the effect of randomness is
negligible is that the tunneling-induced splitting, $t$, of the
states in the layers is the smallest scale in the problem. In fact,
we had  used this condition by restricting the calculations to the
lowest order in $t$. Under this condition, the under-barrier
scattering of the tunneling electron by donors in the classically
forbidden region is exponentially suppressed as compared to the
situation when donors are located in the vicinity of the layers.

\begin{acknowledgments}

This work was supported by DOE, Office of Basic Energy Sciences,
Award No.~DE-FG02-06ER46313 and by NSF under Grant No.~DMR-0503172.
E.M. acknowledges  hospitality of the Kavli Institute for
Theoretical Physics at UCSB and support by the National Science
Foundation under Grant No. PHY99-07949.
\end{acknowledgments}

\end{document}